\documentclass[a4paper,10pt]{article}

\usepackage{publication}
\usepackage{microtype}
\usepackage[style=numeric, sorting=none]{biblatex}
\usepackage{physics}
\usepackage{amssymb}
\usepackage{graphicx} 
\graphicspath{ {figures/} }
\usepackage{caption}
\title{Fast Generation of Metrologically Relevant Fock State Mixtures}

\author[1*]{Gonzalo Reina Rivero}
\contact{* gonzalo.reina@upct.es}
\affil[1]{Universidad Politécnica de Cartagena member of European University of Technology EUT+, Research Group of Quantum Technologies, Área de Física Aplicada, Cartagena E-30202, Spain}

\author[1]{Marcel Morillas-Rozas}

\author[1]{Alberto López-García}

\author[1*]{Javier Cerrillo}
\contact{javier.cerrillo@upct.es}

\date{\today}

\begin{document}

\maketitle

\begin{abstract}
We propose a fast laser pulse sequence for the generation of non-thermal Fock state mixtures of the motion of a trapped ion, targeted at displacement metrology beyond the standard quantum limit. Using a polaron-frame description of the ion–laser interaction, we identify a resonant operating point—zero detuning and a Rabi frequency matching the trap frequency—at which selective population trapping survives strong driving, enabling preparation speeds beyond the weak-driving limit of previous protocols without requiring ground-state cooling. We trace the residual infidelity at large Lamb–Dicke parameter $\eta$ to a single coherent process, the counter-rotating blue-sideband term neglected in the rotating-wave approximation, and show that it is suppressed by two routine calibrations: a percent-level refocusing of the pulse duration and a small compensating Bloch–Siegert detuning. Numerical simulations of the full sequence show that this refinement keeps the preparation error at or below the $10\%$ level up to $\eta\approx0.5$ and restores the displacement-sensing Fisher information that the uncorrected protocol loses at strong coupling, recovering up to 9 dB relative to the nominal sequence.
\end{abstract}


\begin{multicols}{2}

\section{Introduction}

Trapped ions constitute one of the most advanced platforms for quantum information processing \cite{ciracQuantumComputationsCold1995, sorensenQuantumComputationIons1999} and high-precision quantum metrology \cite{winelandQuantumInformationProcessing2011}. The latter branch most notably includes trapped-ion optical clocks and related time-and-frequency experiments, providing highly precise timing standards \cite{ludlowOpticalAtomicClocks2015}. Trapped ions can also serve as ultrasensitive mechanical probes, where metrological gain is extracted from changes in the motional displacement amplitude \cite{wolfMotionalFockStates2019, ivanovQuantumSensingPhasespacedisplacement2018, shanivQuantumLockinForce2017}.

Extremely weak external forces or fields can be inferred from such measurements \cite{UltrasensitiveDetectionForce, ivanovHighprecisionForceSensing2016, QuantumLockinForce}. These schemes underpin, for instance, mass spectroscopy and precision spectroscopy of highly charged ions---``dark'' species lacking convenient laser-cooling transitions (e.g.\ $Ar^{13+}$)---via coupling to a co-trapped ion \cite{kingOpticalAtomicClock2022, leopoldCryogenicRadiofrequencyIon2019}. Displacement sensitivity can also be used to probe electric-field noise near trap electrodes (e.g.\ surface contaminants and patch potentials), a leading contributor to anomalous motional heating in ion-trap quantum processors \cite{mckayMeasurementElectricfieldNoise2021, sedlacekDistanceScalingElectricfield2018}. More broadly, charge-detection mass spectrometry has recently demonstrated the ability to weigh very large particles, including megadalton-scale pathogens \cite{jarroldApplicationsChargeDetection2022}. Motivated by these applications, this work develops a new method for setting up states aimed at metrology that employs displacement-amplitude measurements.

A fundamental challenge in ion-based force and displacement sensing is the standard quantum limit (SQL), which imposes a floor on measurement sensitivity due to the non-vanishing quantum fluctuations of the motional ground state. While non-classical states such as squeezed states \cite{ciracDarkSqueezedStates1993, heinzenQuantumlimitedCoolingDetection1990} and Schrödinger cat states have been successfully implemented to surpass this limit, they often require complex preparation sequences and stringent phase control relative to the measured signal \cite{gilmoreQuantumenhancedSensingDisplacements2021}. Therefore, metrology with squeezed states is best employed to measure displacements inducing a shift perpendicular to the squeezed quadrature.

The excited motional Fock states offer a robust alternative to metrology, providing enhanced sensitivity that scales with the phonon number $n_{ph}$ without the phase-sensitivity constraints of the squeezed states \cite{ciracDarkSqueezedStates1993}. However, the deterministic production of high-$n_{ph}$ Fock states is experimentally demanding, typically necessitating initial ground-state cooling followed by precise blue-sideband pulses on electronic transitions—that require adjusting of laser power to avoid unwanted leakage to neighbouring states.

In our previous work \cite{delakourasProductionFockMixtures2023} we introduced the ``Selective Population Trapping'' (SPT) protocol to generate non-thermal Fock state mixtures, termed trapped states. The method relies on the cyclic application of weak red-sideband (RSB) pulses and electronic spin resets to condense the population into specific motional levels where the RSB-mediated population transfer vanishes; a key advantage is that no initial sideband cooling is required, so metrologically useful states are generated directly from a thermal distribution. That protocol, however, suffers from two intertwined limitations. First, it is fundamentally restricted to the weak-driving regime $\Omega\ll\nu$, both because its effective dynamics were derived perturbatively in this limit and because a weak drive is needed to suppress off-resonant carrier transitions that would otherwise destroy the trapping mechanism; since the trapping rate scales as $\eta\Omega$, this severely caps the achievable preparation speed. Second, the underlying Lamb--Dicke expansion is only valid for $\eta\sqrt{n}\ll1$, confining the scheme to small Lamb--Dicke parameters and low-lying Fock states.

In this work we overcome both limitations. By describing the system in the polaron frame---a treatment that is exact in $\eta$---we identify a specific resonance condition, zero detuning ($\Delta=0$) and a Rabi frequency matching the motional frequency ($\Omega\simeq\nu$), at which the trapping mechanism becomes robust against carrier errors and operates at maximal coupling strength. We further introduce a two-parameter refinement of the driving pulse, based on blue-sideband refocusing and Bloch--Siegert detuning, that maintains the preparation fidelity up to Lamb--Dicke parameters as large as $\eta\approx0.5$. We demonstrate these claims via numerical simulations, benchmarked directly against our original weak-driving protocol at increasing laser intensities.

\section{Theory}

We consider a trapped-ion system driven by a laser field of frequency $\omega_{L}$. The internal degrees of freedom are modelled as a two-level system with a ground state $\ket{\downarrow}$ and an excited state $\ket{\uparrow}$, with a transition frequency $\omega$. The ion's motion is treated as a quantum harmonic oscillator with a natural frequency $\nu$ and the usual ladder operators $a$ and $a^{\dagger}$. In the frame rotating at the laser frequency, obtained through a rotating-wave approximation (RWA) with respect to the optical frequency term ($\omega_{L}\sigma_{z}/2$), the Hamiltonian reads
\begin{equation}
H = \frac{\Delta}{2}\sigma_{z} + \nu a^{\dagger}a + \frac{\Omega}{2}\left[\sigma^{+}D(i\eta) + \sigma^{-}D^{\dagger}(i\eta)\right],
\label{eq:H-start}
\end{equation}
where $\Delta=\omega-\omega_L$ is the laser detuning, $\Omega$ is the Rabi frequency, $\eta$ is the Lamb--Dicke parameter and $D(i\eta)=\exp\!\big[i\eta(a+a^{\dagger})\big]$ is the displacement operator. Since every pulse of the protocol is defined, applied and simulated at this level of description, we refer to Eq.~\eqref{eq:H-start} as the \emph{lab-frame} Hamiltonian in what follows.

The difficulty in treating Eq.~\eqref{eq:H-start} at strong driving resides in the displacement operator attached to the spin-flip terms. The conventional route expands $D(i\eta)$ in powers of $\eta$ and removes rapidly oscillating contributions; this is controlled only when both $\eta\sqrt{n}\ll1$, so that higher Lamb--Dicke orders are negligible, and $\Omega\ll\nu$, so that the off-resonant carrier remains energetically suppressed. At the operating point of interest here---strong driving $\Omega\simeq\nu$ and $\eta$ beyond the Lamb--Dicke regime---both conditions fail simultaneously, and no truncation of the expansion is justified. The polaron transformation resolves this problem exactly rather than perturbatively: we move to the \emph{polaron frame} via the unitary
\begin{equation}
U_{P}=\exp\!\left[\frac{-i\eta}{2}\sigma_{z}\left(a+a^{\dagger}\right)\right],
\label{eq:polaron-U}
\end{equation}
a spin-conditioned displacement chosen precisely so that it cancels the displacement operators of the coupling, $U_{P}\,\sigma^{+}D(i\eta)\,U_{P}^{\dagger}=\sigma^{+}$. The transformed polaron-frame Hamiltonian $H_{P}=U_{P}HU_{P}^{\dagger}$ reads
\begin{equation}
H_{P} = \frac{\Delta}{2}\sigma_{z} + \nu a^{\dagger}a + \frac{i\eta\nu}{2}\sigma_{z}\left(a^{\dagger}-a\right) + \frac{\Omega}{2}\sigma_{x} + \nu\frac{\eta^2}{4},
\label{eq:H-polaron}
\end{equation}
where the drive now enters through the bare operator $\sigma_{x}$ and the entire $\eta$ dependence has been relocated into a spin-dependent force that is linear in the oscillator quadratures. No expansion in $\eta$ is involved: Eq.~\eqref{eq:H-polaron} is exact for arbitrary $\eta$ and phonon number, so the single rotating-wave step performed below remains the only approximation of the derivation. This exactness is what allows the protocol to operate outside the Lamb--Dicke regime, where the perturbative route is unavailable.

To diagonalize the drive term we perform a $\pi/2$ rotation about the $y$ axis, $R_{y}=\exp(+i\frac{\pi}{4}\sigma_{y})$, which maps $\sigma_{z}\to-\sigma_{x}$ and $\sigma_{x}\to\sigma_{z}$. Focusing on resonant driving ($\Delta=0$) and dropping the constant $\nu\eta^{2}/4$, the Hamiltonian in this \emph{rotated polaron frame} becomes
\begin{equation}
\tilde{H} = \nu a^{\dagger}a - \frac{i\eta\nu}{2}\sigma_{x}(a^{\dagger}-a) + \frac{\Omega}{2}\sigma_{z}.
\label{eq:H-final}
\end{equation}
Finally, we move to the interaction picture with respect to the free Hamiltonian $H_{0}=\nu a^{\dagger}a + \frac{\Omega}{2}\sigma_z$---the \emph{effective frame}, in which the trapping dynamics of this work is formulated---and apply a single RWA, retaining only the slowly varying contributions. For the operating point $\Omega = \nu$ this yields the effective Hamiltonian
\begin{equation}
H_{JC} = \frac{i\eta\Omega}{2}(a\sigma^{+} - a^{\dagger}\sigma^{-}).
\label{eq:H-JC}
\end{equation}
This effective generator is formally the same Jaynes--Cummings (JC) interaction that we obtained in the perturbative Lamb--Dicke regime under weak off-resonant driving in our previous work. The two results are, however, physically distinct. There, the coupling arose from the first order of the expansion of $D(i\eta)$ after two successive rotating-wave approximations, one of which already imposed $\Omega\ll\nu$; its validity was therefore restricted to $\eta\sqrt{n}\ll1$ and to slow preparation. Here, Eq.~\eqref{eq:H-JC} is obtained non-perturbatively at resonance: its validity extends beyond the Lamb--Dicke regime, it is controlled by a single RWA whose leading correction we identify and suppress in Section~3.1, and it operates at the maximal coupling $\Omega=\nu$---the origin of the speed-up demonstrated below.

The dynamics generated by $H_{JC}$ fix the duration of the driving pulses of the protocol. Restricted to the subspace $\{\ket{\downarrow,n},\ket{\uparrow,n-1}\}$, Eq.~\eqref{eq:H-JC} induces Rabi oscillations with a generalized Rabi frequency $\Omega_n = \eta \Omega \sqrt{n}$. Throughout this work, the pulse duration $\tau$ is defined as the time of one complete Rabi cycle (a full $2\pi$ rotation) for a target phonon number $n_{0}$,
\begin{equation}
\tau \equiv \frac{2\pi}{\eta \Omega \sqrt{n_0}},
\label{eq:tau}
\end{equation}
so that during one pulse a Fock state $\ket{n}$ is rotated by the angle $2\pi\sqrt{n/n_{0}}$. The inverse $\sqrt{n_0}$ scaling reflects the bosonic enhancement of the coupling strength and implies faster population transfer for larger motional excitations.

This dynamical picture also reveals a significant advantage in the mean trapping rate. Since Eq.~\eqref{eq:H-JC} coincides formally with the effective Hamiltonian of our previous work, the same expression applies \cite{delakourasProductionFockMixtures2023}: for a given Fock state $n$,
\begin{equation}
    \bar{R}_{tr} = \frac{\pi}{\sqrt{n_0}}\mathrm{sinc}^{2}\bigg(\sqrt{\frac{n}{n_0}\pi}\bigg)\frac{\eta\Omega}{2},
\end{equation}
where the mean trapping rate $\bar{R}_{tr}$ depends linearly on both $\eta$ and $\Omega$. Operating at $\Omega = \nu$ therefore increases the trapping rate by orders of magnitude with respect to the weak-driving requirement $\Omega \ll \nu$.

While the polaron transformation provides a compact and non-perturbative derivation of the effective Hamiltonian, the same resonant JC interaction can in principle be obtained directly from the lab-frame Hamiltonian~\eqref{eq:H-start} by setting $\Delta=0$, moving into the strong-driving interaction picture, and performing a RWA at the operating point $\Omega = \nu$. The transformation thus serves as a transparent and systematic route to the same effective description: it makes the exactness in $\eta$ manifest at every step, and it organizes the error analysis of Section~3.1, where the leading correction is identified with a single counter-rotating term of Eq.~\eqref{eq:H-final}.

\section{Numerical Results}

We simulated the protocol using the QuTiP library in Python \cite{lambertQuTiP5Quantum2025}. The dissipative dynamics are modelled with the Lindblad master equation,
\begin{equation}
\frac{d}{dt} \rho = -i [H, \rho] + \frac{\Gamma}{2} \left(2 \sigma^{-} \tilde{\rho} \sigma^{+} - \sigma^{+} \sigma^{-} \rho - \rho \sigma^{+} \sigma^{-}\right)
\equiv \mathcal{L}(\Omega, \Gamma)\rho,
\label{eq:lindblad-master}
\end{equation}
where $\Gamma$ is the decay rate and $\mathcal{L}$ is the Liouvillian superoperator \cite{dowlingExploringQuantumAtoms2014, lindbladGeneratorsQuantumDynamical1976}. Note that the decay term includes a displaced density matrix $\tilde{\rho}$: this accounts for the recoil effect of spontaneous emission, which is described as
\begin{equation}
    \tilde{\rho} = \frac{1}{2} \int^{+1}_{-1}{W(x)e^{ix\eta'(a+a^{\dagger})}\rho e^{-ix\eta'(a+a^{\dagger})}dx},
\end{equation}
where $W(x)$ is a function describing the distribution of angles of spontaneous emission, defined as $W(x)=3(1+x^2)/4$ and $\eta'$ is approximated to be $\eta' \simeq \eta$. The initial state is set to be a thermal distribution at temperature $T$,
\begin{equation}
\rho(0)=\ket{g}\!\bra{g}\otimes \rho_{\mathrm{th}},\qquad
\rho_{\mathrm{th}}=\frac{e^{-\nu a^{\dagger}a/(k_{B}T)}}{Z},
\label{eq:rho-thermal}
\end{equation}
with $Z=\mathrm{Tr}\!\left[e^{-\nu a^{\dagger}a/(k_{B}T)}\right]$.

\begin{center}
 \includegraphics[width=0.45\textwidth]{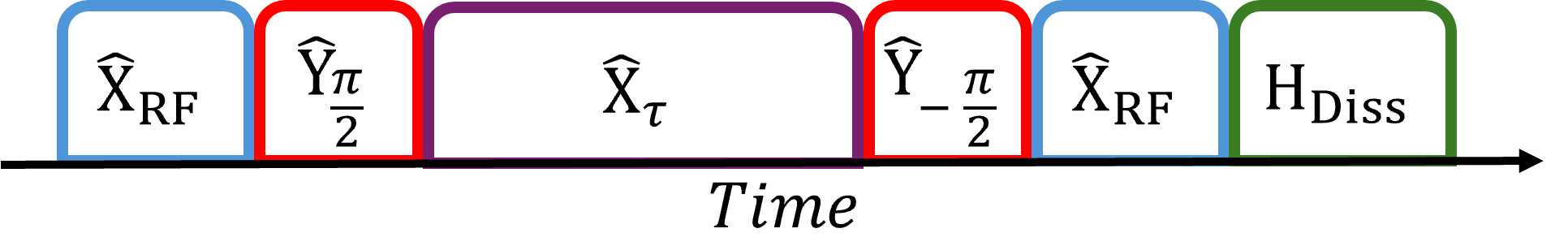}
 \captionof{figure}{Pulse sequence diagram}
 \label{fig:diagram}
\end{center}

To realize the effective dynamics of Eq.~\eqref{eq:H-JC} we implement, within each cycle, the composite pulse sequence illustrated in Figure \ref{fig:diagram}. All pulse Hamiltonians below are written in the lab frame of Eq.~\eqref{eq:H-start}, which is the level at which the time evolution is simulated.
\begin{enumerate}
\item \textit{RF dipolar displacement.} A short motional kick generated by
\begin{equation}
H_{\mathrm{RF}} = \frac{\Omega_{\mathrm{RF}}}{2}\left(a+a^{\dagger}\right),
\end{equation}
applied for a duration $t_{\mathrm{RF}}=\eta/\Omega_{\mathrm{RF}}$.
\item \textit{$Y(\pi/2)$ spin rotation.} A resonant pulse of strength $\Omega_{Y}$ implementing a $\pi/2$ rotation around $y$,
\begin{equation}
H_{Y} = \frac{\Delta}{2}\sigma_{z} + \nu a^{\dagger}a + \frac{i\Omega_{Y}}{2}\left[\sigma^{+}D(i\eta) - \sigma^{-}D^{\dagger}(i\eta)\right],
\end{equation}
applied for $t_{y}=\pi/(2\Omega_{Y})$. In the simulations we fix $\Omega_{Y}=100\nu$.
\item \textit{$X$ pulse.} A strong interaction pulse described by
\begin{equation}
H_{X} = \frac{\Delta}{2}\sigma_{z} + \nu a^{\dagger}a + \frac{\Omega}{2}\left[\sigma^{+}D(i\eta) + \sigma^{-}D^{\dagger}(i\eta)\right],
\end{equation}
applied for the duration $\tau$ of Eq.~\eqref{eq:tau}, where $n_0$ selects the trapped manifold; hereafter we set $n_{0}=1$.
\item \textit{Undo rotations and displacement.} We apply the inverse operations $Y(-\pi/2)$ and $-H_{\mathrm{RF}}$ for the same durations as steps (i) and (ii).
\item \textit{Spin reset.} Finally, we include dissipation that resets the spin back to $\ket{g}$ at the end of each cycle.
\end{enumerate}

This pulse train mirrors the chain of frame transformations of Section~2. The trapping dynamics of Eq.~\eqref{eq:H-JC} are defined in the effective frame, whereas laboratory pulses necessarily act in the lab frame; steps (i) and (ii) bridge the two by actively synthesizing the transformations of the theory: the RF kick imprints the motional displacement of the polaron transformation $U_{P}$ on the freshly reset spin state, and the $Y(\pi/2)$ rotation implements $R_{y}$. The $X$ pulse (iii), a plain resonant drive of duration $\tau$, then realizes the effective JC dynamics: each Fock state $\ket{n}$ is rotated by the angle $2\pi\sqrt{n/n_{0}}$, so the trap states $n=n_{0}m^{2}$ ($m$ integer) complete an integer number of Rabi cycles and are returned intact, whereas all other states are partially transferred towards lower-lying levels. Step (iv) undoes the frame synthesis, returning the state to the lab frame, and the spin reset (v) removes the entropy associated with the transfer, rendering it irreversible. Iterating the cycle therefore funnels the initially thermal population into the trapped manifold.

Throughout this section we employ resonant driving ($\Delta=0$) and the operating point $\Omega \simeq \nu$, effectively replicating the dynamics given by Eq.~\eqref{eq:H-final}. In the simulations, the dynamics of each pulse are computed by constructing the corresponding unitary propagator for each step of the pulse sequence and applying them to the initial state.This yields the stroboscopic evolution for one full cycle, which is iterated and applied to the initial state $\rho(0)$. Unless otherwise stated, all simulations used 30 cycles and a motional state truncation of $N=14$ by default. For reference, the parameters used were $\eta=0.05$, $\Omega=\nu$, $\Omega_{Y}=100\nu$, $\Omega_{\mathrm{RF}}=\nu$, $\Gamma=1000\nu$, $k_{B}T=100\nu$ and $n_0=1$.

\begin{center}
\includegraphics[width=0.45\textwidth]{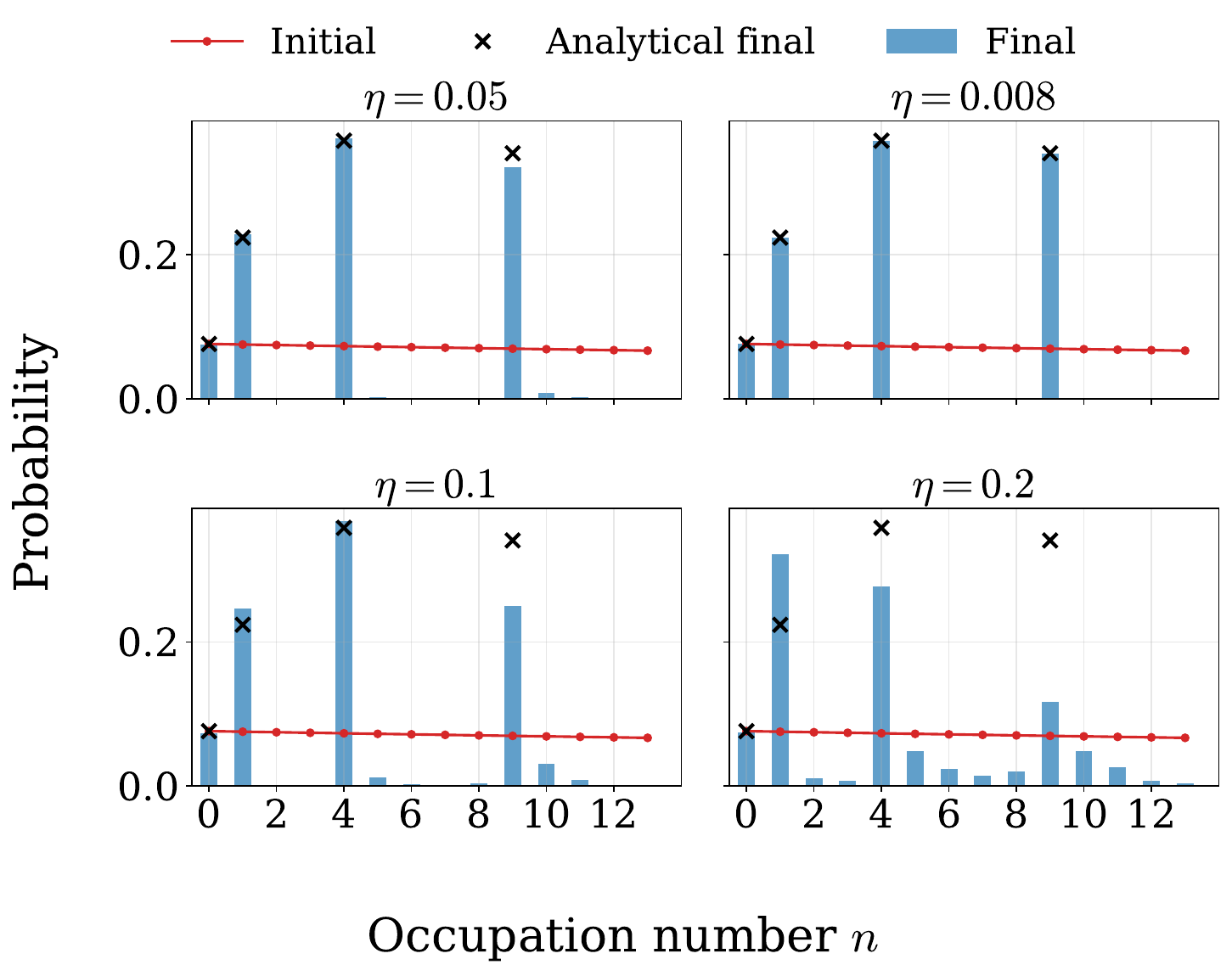}
\captionof{figure}{Final trapped-state populations for increasing values of $\eta$ after 30 cycles of the pulse sequence.}
\label{fig:eta-scan}
\end{center}

Figure~\ref{fig:eta-scan} shows the final motional-state populations for several values of the Lamb--Dicke parameter $\eta$. We observe that increasing $\eta$ degrades the quality of trapping for higher-lying motional states. This behaviour is consistent with the growing importance of counter-rotating contributions in the effective JC-like description, which become more relevant as the coupling strength is increased \cite{lizuainVibrationalBlochSiegertEffect2008}. To quantify this behaviour and compare directly against the previous SPT sequence, we use the trace distance between the final motional state and the ideal trapped distribution,
\begin{equation}
    p_{tr}(m)=\sum^{n_{0}(m+1)^2-1}_{k=n_{0}m^2}{\bra{k}\mu_0\ket{k}}
\end{equation}
where $\mu_0$ is the initial motional state, as introduced in our previous work \cite{delakourasProductionFockMixtures2023}. This expression essentially traps any initial states below $\ket{n_0(m+1)^2}$ into trap state $\ket{n_0m^2}$.

Figure~\ref{fig:omega-scan} summarizes the main comparison between protocols. Panel A) reports the trace distance for the legacy SPT sequence over a range of driving strengths $\Omega$ and for several values of $\eta$, together with the value obtained with the new resonant sequence at its fixed operating point $\Omega=\nu$. Even though the new protocol uses only this single resonant value of $\Omega$, it still yields a smaller trace distance than the previous SPT protocol at the same driving strength, including for the largest explored $\eta$ values.

Panel B) complements this comparison by plotting the trace distance as a function of $\eta$. The new sequence remains closer to the ideal trapped state over a broad interval and outperforms the previous protocol up to approximately $\eta\approx0.4$, where both approaches tend to converge. This identifies the practical robustness window of the resonant implementation as the Lamb--Dicke parameter is increased.

We now observe the resulting speed-up. Figure~\ref{fig:convergence} tracks the trace distance to the ideal trapped distribution as a function of the elapsed preparation time, comparing the resonant sequence ($\Omega=\nu$) with the weak-driving SPT protocol at $\Omega=\nu/100$ and $\Omega=\nu/1000$, all at the reference value $\eta=0.05$. The resonant sequence funnels the population into the trapped manifold within about $10$ cycles, reaching a trace distance of $2\times10^{-2}$ at a preparation time $t\nu\sim10^{3}$. Because a single Rabi cycle lasts $\tau\propto1/\Omega$, lowering the drive by one and two orders of magnitude shifts the corresponding curves rigidly to the right by the same factors, so that the weak-driving protocol requires $10^{2}$--$10^{3}$ times longer to reach a comparable---indeed slightly worse---trapping quality. This is precisely the linear scaling of the mean trapping rate $\bar{R}_{tr}\propto\eta\Omega$: operating at the maximal resonant coupling $\Omega=\nu$ minimizes the preparation time at fixed cycle count, while the non-perturbative construction additionally lowers the residual error floor relative to the weak-driving limit.

\begin{center}
\includegraphics[width=0.45\textwidth]{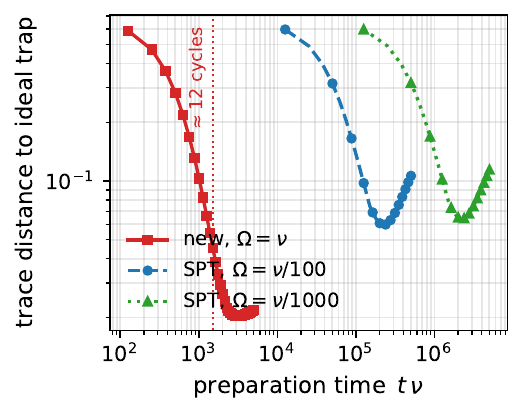}
\captionof{figure}{Preparation speed at $\eta=0.05$. Trace distance to the ideal trapped distribution versus elapsed preparation time $t\nu$ for the resonant sequence ($\Omega=\nu$) and the weak-driving SPT protocol ($\Omega=\nu/100$ and $\nu/1000$); markers denote successive cycles. The resonant sequence saturates within about $10$ cycles (dotted line), whereas reducing $\Omega$ shifts the curves rigidly towards longer times, reflecting the linear trapping-rate scaling $\bar{R}_{tr}\propto\eta\Omega$.}
\label{fig:convergence}
\end{center}

\begin{center}
 \centering
 \includegraphics[width=0.45\textwidth]{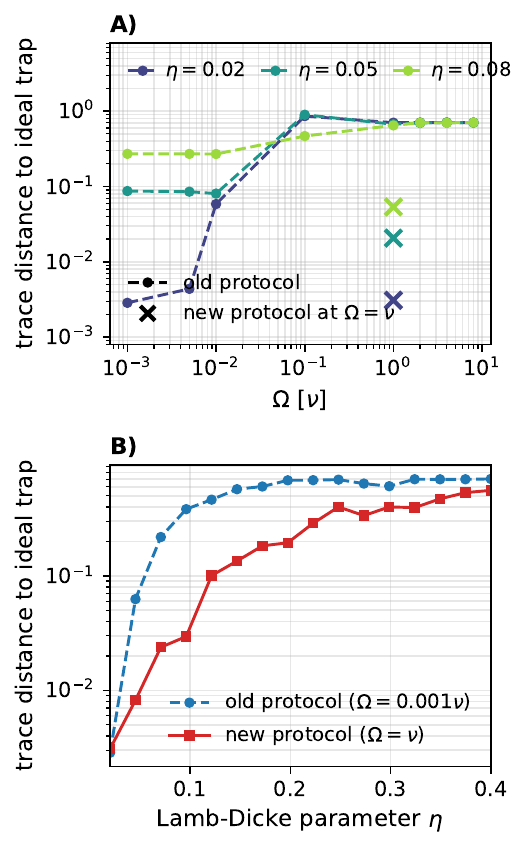}
 \captionof{figure}{Comparison between the legacy SPT and the new resonant sequence. A) Trace distance to the ideal trapped state as a function of the driving strength $\Omega$ for several $\eta$ in the previous SPT protocol (dashed curves), together with the new protocol evaluated at its fixed operating point $\Omega=\nu$ (crosses). B) Trace distance versus $\eta$, showing improved trapping for the new protocol up to about $\eta\approx0.4$.} 
 \label{fig:omega-scan}
\end{center}

These results indicate that the resonant operating point preserves the selective trapping mechanism while improving preparation speed in the experimentally relevant parameter range.

\subsection{Blue-sideband refocusing with Bloch–Siegert detuning}

Despite the improvement in preparation speed, the proposed protocol still suffers from population leakage at large $\eta$. This leakage can, however, be traced back to a single, well-identified coherent process, and suppressed by a suitable adjustment of the pulse parameters.

The origin of the leakage becomes transparent in the rotated polaron frame. Equation~\eqref{eq:H-final} is the quantum Rabi model with qubit splitting $\Omega$, oscillator frequency $\nu$ and coupling strength $\eta\nu/2$. In the interaction picture with respect to $H_{0}$ at the operating point $\Omega=\nu$, the coupling splits into the resonant JC term of Eq.~\eqref{eq:H-JC}, which generates the trapping dynamics, and a counter-rotating (anti-JC) contribution, $-\frac{i\eta\nu}{2}(\sigma^{+}a^{\dagger}e^{2i\nu t}-\mathrm{h.c.})$, oscillating at $2\nu$, which is discarded by the RWA. The latter drives off-resonant blue-sideband excursions $\ket{\downarrow,n}\to\ket{\uparrow,n+1}$ with matrix element $\eta\nu\sqrt{n+1}/2$, so the accuracy of the RWA is governed by the ratio $\eta\sqrt{n+1}/4$: the approximation---and with it the trapping---degrades precisely when $\eta$ is no longer small \cite{lvQuantumSimulationQuantum2018}. A controlled numerical test confirms this picture: removing the counter-rotating term by hand from the otherwise exact coherent dynamics restores trapping to machine precision for all $\eta$ considered, identifying the blue sideband as the sole coherent leakage channel. This is the vibrational analogue of the Bloch--Siegert regime of strongly driven two-level systems \cite{blochMagneticResonanceNonrotating1940, lizuainVibrationalBlochSiegertEffect2008}.

This diagnosis suggests a remedy that requires neither new hardware nor a departure from the resonant operating point. Because the blue-sideband excursion is coherent and detuned by $2\nu$, it returns to the trap state whenever the generalized Rabi phase accumulated over the pulse is a multiple of $2\pi$, i.e.\ $\Omega_{g}(n)\tau\in2\pi\mathbb{Z}$ with $\Omega_{g}(n)=\sqrt{(2\nu)^{2}+\eta^{2}\nu^{2}(n+1)}$. Since $\Omega_{g}$ depends only weakly on $n$, a small rescaling of the pulse duration, $\tau\to f\tau$ with $f$ within a few percent of unity, collectively refocuses the excursions of all relevant trap states. The residual effect is a coherent level shift of order $\eta^{2}\nu/8$---the Bloch--Siegert shift---which is compensated by adding a small static detuning $\delta$ to the $X$ pulse. Both corrections involve only parameters that are calibrated routinely in trapped-ion experiments: choosing the drive detuning and pulse duration so that off-resonant excursions close at the end of the pulse is standard practice in trapped-ion entangling gates \cite{sorensenEntanglementQuantumComputation2000, leibfriedExperimentalDemonstrationRobust2003, leungRobust2QubitGates2018, ballanceHighFidelityQuantumLogic2016, gaeblerHighFidelityUniversalGate2016, clarkHighFidelityBellStatePreparation2021}, and vibrational Bloch--Siegert shifts of the sideband resonances can be resolved and compensated experimentally \cite{lizuainVibrationalBlochSiegertEffect2008, haffnerPrecisionMeasurementCompensation2003}.

\begin{center}
\includegraphics[width=0.45\textwidth]{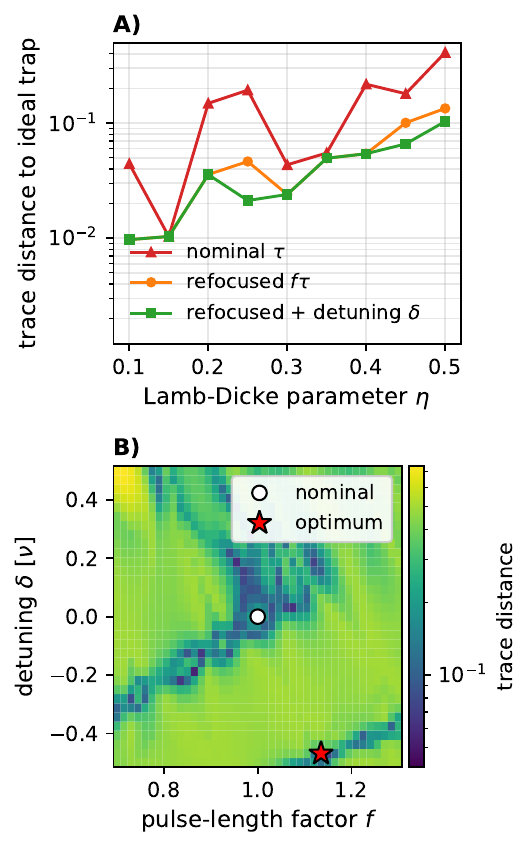}
\captionof{figure}{Blue-sideband refocusing and Bloch--Siegert detuning on the full pulse sequence. A) Trace distance to the ideal trapped distribution versus $\eta$ for the nominal pulse duration $\tau$, the refocused duration $f\tau$, and the refocused duration combined with a compensating detuning $\delta$. B) Calibration landscape at $\eta=0.4$: trace distance as a function of the pulse-length factor $f$ and the detuning $\delta$, with the nominal operating point (circle) and the optimum (star) indicated.}
\label{fig:refocus}
\end{center}

Figure~\ref{fig:refocus}A) validates the refinement on the full pulse sequence of Figure~\ref{fig:diagram}. These and the following simulations use a motional truncation of $N=16$, an initial thermal distribution with mean occupation $\bar{n}=1$, $40$ cycles, and the full dissipative reset model of Eq.~\eqref{eq:lindblad-master}, including photon recoil ($N=24$ and $\bar{n}=5$ for the metrological benchmark below). Refocusing the pulse duration alone reduces the trace distance to the ideal trapped distribution by up to an order of magnitude (from $0.15$ to $0.04$ at $\eta=0.2$, and from $0.22$ to $0.05$ at $\eta=0.4$), and the optimal $f$ remains within $5\%$ of unity throughout. The compensating detuning---whose optimal magnitude grows with $\eta$, as expected for an $\eta^{2}$ effect---provides additional margin for $\eta\gtrsim0.4$, keeping the preparation error at the few-percent level up to $\eta=0.45$ and reducing it from $0.41$ to $0.10$ at $\eta=0.5$. The non-monotonic behaviour of the nominal protocol is itself a signature of the refocusing mechanism: particularly favourable $\eta$ values are those for which the nominal $\tau$ happens to be nearly commensurate with the $2\nu$ oscillation. Figure~\ref{fig:refocus}B) shows the corresponding two-parameter calibration landscape at $\eta=0.4$, where already the small correction ($f=0.96$, $\delta=0$) lowers the trace distance from $0.22$ to $0.05$, while the global optimum of the scanned window lies on a neighbouring refocusing branch ($f=1.14$, $\delta=-0.47\nu$) and reaches $0.04$. This map can be obtained experimentally through a standard two-parameter calibration scan of the pulse duration and detuning. We have verified that an idealized simulation with instantaneous, recoil-free spin reset yields nearly identical optima, confirming that the calibration is fixed by the coherent blue-sideband dynamics rather than by the reset.

\begin{center}
\includegraphics[width=0.45\textwidth]{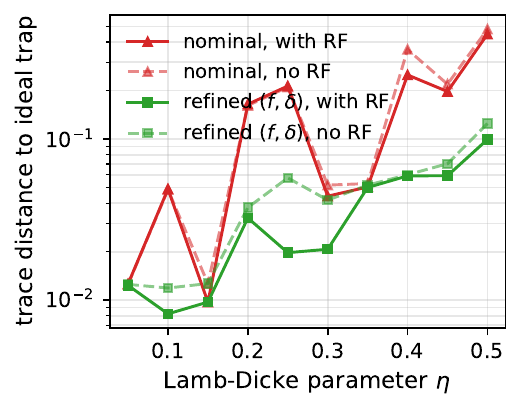}
\captionof{figure}{Contribution of the RF displacement pulse to the preparation fidelity. Trace distance to the ideal trapped distribution for the nominal and the refined (refocused and detuned) sequences, with and without the RF pulse; for the refined curves, $(f,\delta)$ are re-calibrated separately for each variant.}
\label{fig:rf}
\end{center}

The same analysis clarifies the role of the RF displacement pulse in the sequence. This kick realizes the motional displacement of the polaron transformation on the freshly reset spin state, so, although its amplitude $\eta/2$ is small, its contribution grows with the coupling strength. Figure~\ref{fig:rf} compares the nominal and refined sequences with and without it: for $\eta\lesssim0.35$ the two variants are nearly indistinguishable, whereas at stronger coupling omitting the pulse increases the preparation error of the nominal sequence from $0.22$ to $0.32$ at $\eta=0.4$, and that of the refined sequence---even after re-calibrating $(f,\delta)$ for the modified sequence---from $0.10$ to $0.13$ at $\eta=0.5$. Since the pulse adds only a duration $\eta/\Omega_{\mathrm{RF}}$ per cycle, it is retained throughout.

\begin{center}
\includegraphics[width=0.45\textwidth]{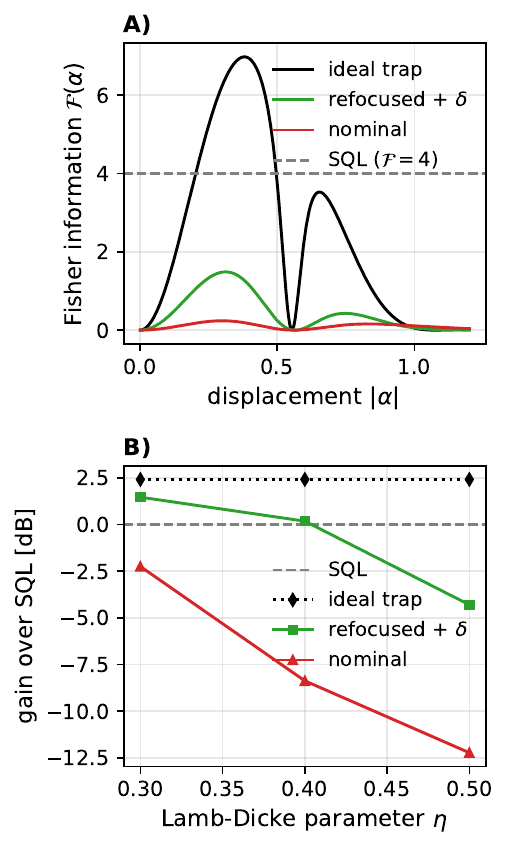}
\captionof{figure}{Metrological benchmark of the refined preparation for an initial thermal state with $\bar{n}=5$ and readout of the Fock state $\ket{n=4}$. A) Fisher information of a phase-space displacement $\alpha$ at $\eta=0.5$ for the ideal trapped mixture, the refined (refocused and detuned) preparation, and the nominal protocol; the dashed line marks the SQL, $\mathcal{F}=4$. B) Gain over the SQL as a function of $\eta$.}
\label{fig:metrology}
\end{center}

The refinement is not merely a matter of state fidelity: it restores the metrological advantage that motivates the protocol. Following the displacement-sensing analysis of our previous work \cite{delakourasProductionFockMixtures2023}, we evaluate the classical Fisher information of a phase-space displacement $\alpha$ read out through the population of a single Fock state, and compare it with the SQL value $\mathcal{F}_{\mathrm{SQL}}=4$ set by the motional ground state. Starting from a hotter thermal distribution ($\bar{n}=5$), so that the trap concentrates population in the excited state $n=4$ where both the metrological gain and the counter-rotating coupling ($\propto\sqrt{n+1}$) are largest, the ideal trapped mixture provides a gain of $2.4~\mathrm{dB}$ over the SQL independently of $\eta$. As shown in Figure~\ref{fig:metrology}, the nominal protocol loses this advantage entirely at strong coupling, falling $8$--$12~\mathrm{dB}$ below the SQL for $\eta=0.4$--$0.5$, whereas the refined preparation recovers $4$--$9~\mathrm{dB}$ of Fisher information: it remains at or above the SQL up to $\eta=0.4$ and within $5~\mathrm{dB}$ of it at $\eta=0.5$.

\begin{center}
\includegraphics[width=0.45\textwidth]{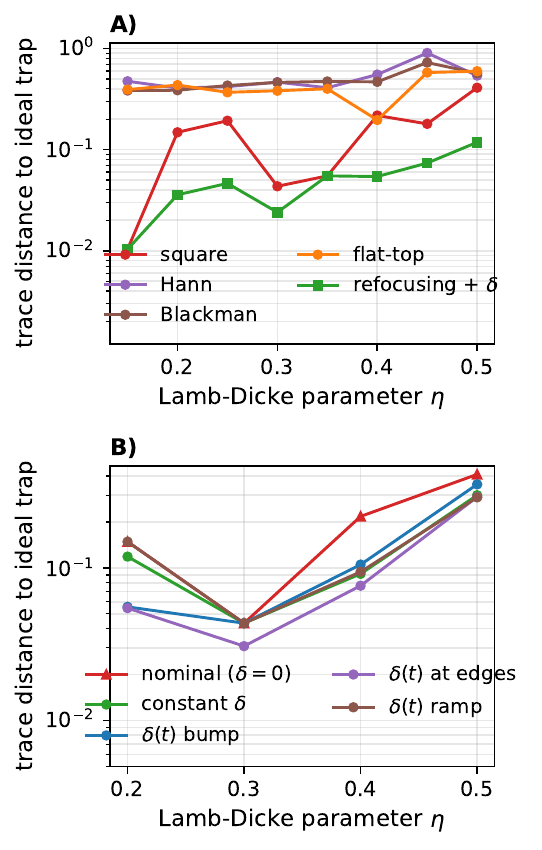}
\captionof{figure}{Pulse shaping at the resonant operating point. A) Trace distance to the ideal trapped distribution for area-matched smooth amplitude envelopes (Hann, Blackman, cosine-ramped flat-top) compared with the square pulse and with refocusing plus detuning. B) Time-dependent detuning profiles $\delta(t)$ (with amplitude optimized at each $\eta$ and $f=1$) compared with the nominal protocol.}
\label{fig:shaping}
\end{center}

Finally, we address whether smooth amplitude shaping of the $X$ pulse---the standard remedy against off-resonant excitation---could achieve the same end \cite{thomAccurateAgileDigital2013, sanerBreakingEntanglingGate2023, zarantonelloRobustResourceEfficientMicrowave2019}. It cannot, because at resonance the Rabi frequency plays a dual role: it sets the drive strength while simultaneously pinning the blue sideband at its maximal detuning $\Omega+\nu=2\nu$. Any smooth envelope necessarily sweeps $\Omega(t)$ below $\nu$ during the ramps, dragging the counter-rotating term toward resonance. Accordingly, area-matched Hann, Blackman and flat-top envelopes perform systematically worse than the plain square pulse, and far worse than refocusing with detuning [Figure~\ref{fig:shaping}A)]; a control simulation on the pure JC dynamics, which depend only on the pulse area, is insensitive to the envelope, confirming the mechanism. The modulation compatible with $\Omega=\nu$ is instead a time-dependent detuning $\delta(t)$, which generalizes the constant Bloch--Siegert compensation while leaving the drive amplitude unaltered. As shown in Figure~\ref{fig:shaping}B), no detuning profile degrades the nominal protocol, and shaped profiles outperform the constant detuning at small and moderate $\eta$ (by roughly a factor of $2$ at $\eta=0.2$). Together with the pulse-length factor $f$, the detuning waveform thus constitutes the natural pulse-shaping space for the resonant protocol.

\section{Conclusions}
We have introduced a resonant pulse-sequence implementation of selective population trapping that operates at strong driving, targeting the regime $\Omega\simeq\nu$ where standard sideband-resolved protocols become challenging. By using a polaron-frame viewpoint, we identified conditions under which carrier-type errors are mitigated and the trapping mechanism survives beyond the weak-driving limit. Numerical simulations show that, starting from a thermal motional distribution, the population can be funnelled into a trapped manifold within a modest number of cycles, while remaining robust over a range of $\eta$ and increasing $\Omega$.

Furthermore, we traced the residual infidelity of the resonant protocol at large $\eta$ to a single coherent process---the counter-rotating blue-sideband term neglected by the RWA---and showed that it can be mitigated by two calibrations that are routine in any ion-trap experiment: a percent-level refocusing of the pulse duration and a small Bloch--Siegert compensating detuning. This refinement keeps the preparation error at the few-percent level up to $\eta\approx0.45$, reaching the $10\%$ level at $\eta=0.5$, and, crucially, restores the displacement-sensing Fisher information that the uncorrected protocol loses at strong coupling, recovering $4$--$9~\mathrm{dB}$ relative to the nominal sequence. We also showed that conventional amplitude pulse shaping is counterproductive at the resonant operating point, and that the appropriate shaped degree of freedom is instead a time-dependent detuning.

These results suggest a practical route to faster preparation of motional nonclassical resources without relying on ground-state cooling or long sequences of narrowband sideband pulses. Natural next steps include extending the sensing benchmark to optimized readout strategies and quantum Fisher information under realistic constraints, as well as incorporating motional heating, laser-amplitude noise, and imperfect spin reset, which will further clarify the operational parameter window for experiments and quantify the trade-off between preparation speed and metrological gain.

\end{multicols}

\printbibliography

\newpage

\begin{appendices}
    \section{Funding}
    Authors acknowledge support from grant CNS2023-144994 funded by MICIU/AEI/10.13039/501100011033 and by “ERDF/EU”. JC acknowledges European
Union project C-QuENS (Grant No. 101135359).
    \section{Data Availability}
    All simulation code and graphs are available on \url{https://github.com/GonReina/MixturesPaper.git}.
\end{appendices}

\end{document}